\input harvmac
\input epsf



\lref\rAbdalla{E. Abdalla and M.C.B. Abdalla, Phys. Lett. B337 (1994) 347;
E. Abdalla and M.C. Abdalla, Int. Jour. Mod. Phys. A10, 
(1995) 1611; E. Abdalla and M.C. Abdalla, Phys. Rev. D52 (1995) 6660;
E. Abdalla and M.C. Abdalla, Phys. Rept. 265 (1996) 253.}
\lref\rKG{D.J. Gross, I.R. Klebanov, A.V. Matytsin and A.V. Smilga,
Nucl. Phys. B461 (1996) 109.}
\lref\rthacker{H.B. Thacker, Rev. Mod. Phys. 53 (1981) 253;
N. Andrei, K. Furuya and J.H. Lowenstein, Rev. Mod. Phys. 55 (1983)
331; A.Zamolodchikov and A.Zamolodchikov, Ann. Phys. 120 (1979) 253;
L. Faddeev, in Les Houches Summer School in Theoretical Physics, 
Session 19:Recent Advances in Field Theory and Statistical Mechanics, 
Les Houches, France; Aug2-Sep10, 1982.}
\lref\rZam{A.Zamolodchikov and A.Zamolodchikov, Ann. Phys. 120 (1979) 253.}
\lref\rStau{W.Krauth and M.Staudacher, Phys. Lett. B388 (1996) 808.}
\lref\rAbdala{E.Abdalla and R.Mohayaee,{\it 'Quasi-integrability and 
two-dimensional $QCD$'}, CERN Preprint CERN-TH/96-253, hep-th/9610059.}
\lref\rUs{R. de Mello Koch and J.P. Rodrigues, {\it 'Classical integrability
of chiral $QCD_{2}$ and classical curves'}, CNLS Preprint CNLS-97-01,
hep-th/9701138, to be published in Mod. Phys. Lett. A.}
\lref\rOrfanidis{S.J.Orfanidis, Phys. Rev. D14 (1976) 472.}
\lref\rBuch{W.B\"uchm\"uller, S.T.Love and R.D. Peccei, 
Phys. Lett. 108B (1982) 426. }
\lref\rNevPap{A. Neveu and N. Papanicoloau, Commun. Math. Phys. 58 
(1978) 31.}
\lref\rPap{N. Papanicoloau, Ann. Phys. 136 (1981) 210.}
\lref\rFaddeev{L. Faddeev, in Les Houches Summer School in Theoretical 
Physics, Session 19:Recent Advances in Field Theory and Statistical 
Mechanics, Les Houches, France, Aug2-Sep10, 1982.}
\lref\rCalini{A.Calini and T.Ivey, {\it 'Backlund transformations and knots of
constant torsion'}, Preprint dg-ga/9608001.}
\lref\rRo{C.Rogers in "Soliton Theory: a survey of results", ed. A.P. Fordy,
St. Martin's Press, 1990.}
\lref\rTak{L. Takhtajan, Phys. Lett. A64 (1977) 235.}
\lref\rFog{H.Fogedby, J. Phys. A Math. Gen. 13 (1980) 1467.}
\lref\rHas{Hasimoto, J. Fluid Mech. 51 (1972) 477.}
\lref\rLamb{G.L. Lamb, Jour. Mth. Phys. 18, (1977) 1654.}
\lref\rSchwarz{J.H. Schwarz, hep-th 9505170.}
\lref\rCoote{C.G. Callan Jr., N. Coote and D.J. Gross, Phys. Rev. D13 (1976)
1649.}
\lref\rTHooft{G.'t Hooft, Nucl. Phys. B75 (1974) 461.}
\lref\rSchon{E. Abdalla, R. Mohayaee and A. Zadra, {\it 'Screening in 
Two-dimensional $QCD$'}, Preprint IC/96/51, hep-th 9604063; R. Mohayee,
{\it 'The phases of two dimensional $QED$ and $QCD$'},
hep-th 9705243; Y. Frishman and J. Sonnenschein, Nucl. Phys. B496 (1997) 285;
J. Sonnenschein, {\it 'More on Screening and Confinement in $2D$ $QCD$'}, 
hep-th 9707031; A. Armoni and J. Sonnenschein, {\it 'Screening and Confinement
in large $N_{f}$ $QCD_{2}$ and in $N=1$ $SYM_{2}$'}, Preprint TAUP-2412-97,
hep-th 9703114.}
\lref\rStelle{E. Cremmer, H. L\"u, C.N. Pope and K.S. Stelle, 
{\it 'Spectrum-generating Symmetries from BPS Solitons'}, Preprint
LTP TAMU-29/97, Imperial/TP/96-97/54, LPTENS-97/34, SISSARef 92/97/EP,
hep-th 9707207.}
\lref\rUsT{R. de Mello Koch and J.P. Rodrigues, Phys. Rev. D54, 7794 (1996).}
\lref\rUst{R. de Mello Koch and J.P. Rodrigues, in preparation.}
\lref\rJevicki{A. Jevicki and N. Papanicoloau, Ann. Phys. 120 (1979) 107.}
\lref\rcc{P.G. Grinevich and M.U. Schmidt, {\it 'Closed Curves in $R^{3}$:
a characterization in terms of curvature and torsion, the Hasimoto map and
periodic solutions to the Filament equation'}, dg-ga 9703020.}
\lref\rMarsh{R. Donagi and E. Witten, Nucl. Phys. B460 (1996) 299.}
\lref\rY{M.U. Schmidt, Memoirs of the Amer. Math. Soc. (1996) 551.}

\Title{CNLS-97-04}
{\vbox {\centerline{The Dynamics of Classical Chiral $QCD_{2}$ Currents}
}}

\smallskip
\centerline{Robert de Mello Koch and Jo\~ao P. Rodrigues}
\smallskip
\centerline{\it Physics Department and Centre for Nonlinear Studies}
\centerline{\it University of the Witwatersrand}
\centerline{\it Wits 2050, South Africa}\bigskip

\medskip

\noindent

In this paper the dynamics of the classical chiral $QCD_{2}$ currents
is studied. We describe how the dynamics of the theory can be
summarized in an equation of the Lax form, thereby demonstrating the
existence of an infinite set of conserved quantities. Next, the $r$ matrix
of a fundamental Poisson relation is obtained and used to demonstrate that
the conserved charges Poisson commute. An underlying diffeomorphism symmetry
of the equations of motion which is not a symmetry of the action
is used to provide a geometric interpretation for
the case of gauge group $SU(2)$. This enables us to show that the solutions
to the classical equations of motion can be identified with a large class
of curves, to demonstrate an auto-B\"acklund transformation and to
demonstrate a non linear superposition principle. 
A link between the spectral problem for $QCD_{2}$ and the solution to the 
closed curve problem is also demonstrated.
We then go on to
provide a systematic inverse scattering treatment. 
This formalism is used
to obtain the reflectionless single boundstate eigenvalue soliton solution.


\Date{August, 1997}


\newsec{Introduction.}

$QCD_{2}$ with massless quarks in the fundamental representation of the gauge
group has received considerable attention recently. This has occurred for
essentially two reasons: $\quad\quad$
the first is the claim using non-abelian bosonization
that the theory should be integrable \rAbdalla\ ; the second stems
from the observation \rKG\ that for massless quarks in a
representation with dimension larger than or equal to the dimension of the
representation of a heavy probe charge, the theory exhibits screening.

In $1+1$ dimensions, integrability is usually associated with the existence of
a large number of soliton-like solutions of the classical equations of motion
which are recast in a Lax form, from which the existence of an infinite
number of conservation laws follows. If these survive quantization, they
normally imply an elastic, computable S-matrix \rthacker\ with no particle
production.

There is, of course, considerable evidence of particle production
\rCoote\ \rStau\
in the large $N_{c}$ expansion of 
$QCD_{2}$ \rTHooft\ . It has been argued recently
that this is not necessarily inconsistent with the existence of an
integrable sector in the theory \rAbdala\ . The issue of screening versus
confinement for massless quarks
in the fundamental representation of SU(N) has also been discussed recently
\rSchon\ .

In a previous communication \rUs\ , we analyzed the classical equations of
motion of the theory in the light cone gauge $A_{-}=0$
and established that the
once integrated equation of motion for the currents is of the Lax form,
demonstrating the existence of a infinite number of conserved quantities.
We discovered that the equations of motion possess a continuous symmetry
(diffeomorphism invariance in the $x^{-}$ coordinate) which is not a symmetry
of the action \foot{Recently, a similar symmetry has been noted in
$IIB$ supergravity\rStelle\ .}.
Making use of this symmetry and specializing to
SU(2), we established that all classical Frenet-Serret curves with a constant
term in the torsion are solutions of the equations of motion, and constructed
a single soliton solution.

The original motivation of \rUs\ was to use these classical solutions to
construct bilinears \rUsT\ relevant to semiclassical approximations of 
fermionic
theories. This has been implemented for the Gross-Neveu model \rPap\ and the
Thirring model \rOrfanidis\ .

In this article, we are only interested in exploiting further the rich
mathematical structure that emerges in the study of the equations of
motion of the classical currents of the model. Specifically, in the gauge
$A_{-}=0$ these take the form

\eqn\FLaxExplicitT
{\partial_{+}\partial_{-}^{2}\sigma ={ig^{2}\over\sqrt{2}}
[\partial_{-}^{2}\sigma ,\sigma].}
                         
where

\eqn\DefineThis
{\partial_{-}^{2} \sigma_{ab} = 2j_{ab},}

and $j_{ab}$ is an SU(N) current\foot{These equations of motion also
arise in certain compactifications of M(atrix) theory\rUst\ .}.

In section 2, we show how equation \FLaxExplicitT\ 
is arrived at and point out its Lax
form. Following Fadeev, we construct the fundamental Poisson relation and
recast the once integrated equation of motion as a zero curvature condition.
In section 3, the chiral diffeomorphism in the 
$x^{-}$ coordinate is exhibited and for $SU(2)$ it is used to map the 
most general solution of the equations of
motion to appropriate Frenet-Serret curves. Using Darboux coordinates a second
zero curvature condition is obtained. The Auto-Backlund transformation is 
identified and is used to obtain the single soliton solution of \rUs\ . This
Auto-Backlund transformation is then used to construct a non-linear
superposition principle. In particular, we show how the full set of soliton
solutions of the Sine-Gordon model are a subset of solutions generated by
this non-linear superposition principle. For these solitons, we show further
the the fermion number is quantized. The spectral problem of the zero
curvature condition is found to be closely related to the spectral problem
used to solve the closed curve problem. In section 4 our single soliton
solution is shown to be the reflectionless single bound state eigenvalue
inverse scattering solution. This construction is straightforwardly
generalizable to $SU(N)$.

\newsec{Some general observations.}

We work in light cone co-ordinates\foot{Our conventions are
$x^{\pm}={1\over\sqrt{2}}(t\pm x)$}, in the axial gauge $A_{-}=0$. In terms
of the quark fields $\psi_{a}=\big[\psi_{-a},\psi_{+a}\big]^{T}$ which are
in the fundamental representation of $SU(N)$ $(a=1,2,...N),$ the Lagrangian
density is given by

\eqn\FrstLag
{\eqalign{
{\cal L}&={1\over 2}Tr(\partial_{-}A_{+})^{2}
+i\sqrt{2}{\psi_{+a}}^{\dagger}(x)\partial_{-}
\psi_{+a}(x)
+\sqrt{2}{\psi_{-a}}^{\dagger}(x)(i\delta_{ab}\partial_{+}
+aA_{+ab})\psi_{-b}(x)\cr
&-m({\psi_{-a}}^{\dagger}\psi_{+a}+{\psi_{+a}}^{\dagger}\psi_{-a}).}}

The equations of motion for both $A_{+}$ and $\psi_{+}$ do not involve
derivatives with respect to $x^{+}$ and thus are interpreted as constraints.
As usual, the constraints may be solved, leaving an effective theory for the
quarks $(x=(x^{+},x^{-}))$:

\eqn\QCDLagrangian{\eqalign{ {\cal L}&= 
i\sqrt{2}{\psi_{a}}^{\dagger}(x)\partial_{+}
\psi_{a}(x)+ {g^{2}\over 2} \int dy^{-}\times\cr
&\times {\psi_{a}}^{\dagger}(x)\psi_{b}
(x) |x^{-}-y^{-}|{\psi_{b}}^{\dagger}
(x^{+},y^{-})\psi_{a}(x^{+},y^{-})-{g^{2}\over 2N}\int dy^{-}\times\cr
&\times{\psi_{a}}^{\dagger}(x)\psi_{a}(x)|x^{-}-y^{-}|
{\psi_{b}}^{\dagger}(x^{+},y^{-})\psi_{b}(x^{+},y^{-}).}}

where we have let $m\to 0$ and $\psi_{-a}\to\psi_{a}.$
The dynamics of the classical system now follows from this Lagrangian in the
usual way

\eqn\EqnMot
{i\partial_{+}\psi_{a}(x)={g^{2}\over\sqrt{2}}\sigma_{ab}(x)\psi_{b}(x) }

where

\eqn\Sigm
{\sigma_{ab}(x)= \int dy^{-}|x^{-}-y^{-}|\big[\psi_{a}(y^{-})
\psi_{b}^{\dagger}(y^{-})-{1\over N}\delta_{ab}\psi_{c}(x^{+},y^{-})
{\psi_{c}}^{\dagger}(x^{+},y^{-})\big].}

In this last expression, all fields are evaluated at the same point in
"time" $x^{+}$. In the above equations, the $\psi$ fields are 
normal commuting functions. Note that from the definition \Sigm\ we
find

\eqn\SigmPsi
{\partial_{-}^{2}\sigma_{ab}(x)=2\Big(\psi_{a}(x)\psi^{\dagger}_{b}(x)
-{1\over N}\delta_{ab}\psi_{c}(x^{+},y^{-})
{\psi_{c}}^{\dagger}(x^{+},y^{-})\Big).}

It is useful to note that the classical 
equations of motion \EqnMot\ and \Sigm\
arise as the classical dynamics of the Hamiltonian

\eqn\Hamiltonian
{H=-{g^{2}\over 4}\int d^{2}xTr(\partial_{-}^{2}\sigma)\sigma,}

together with the (equal $x^{+}$) Poisson brackets

\eqn\PoissonBrackets
{\{\psi_{a}(x),\psi_{b}^{\dagger}(y) \}= {i\over\sqrt{2}}
\delta_{ab}\delta (x-y).}

An explicit realization for the Poisson brackets is

\eqn\PoissonBracketsTwo
{\{ \alpha,\beta \} = {i\over\sqrt{2}}\int dx \sum_{a}
\Big({\delta\alpha\over\delta\psi_{a}(x)}
{\delta\beta\over\delta\psi_{a}^{\dagger}(x)}-
{\delta\beta\over\delta\psi_{a}(x)}
{\delta\alpha\over\delta\psi_{a}^{\dagger}(x)}\Big) .}

\subsec{Lax Representation and Conservation Laws.}

Using the equations of motion, the time
dependence of $\psi_{a}\psi_{b}^{\dagger}$ is computed as

\eqn\FLaxPair
{\partial_{+}(\psi_{a}\psi_{b}^{\dagger})={ig^{2}\over\sqrt{2}}
[\psi_{a}\psi_{c}^{\dagger}\sigma_{cb}-
\sigma_{ac}\psi_{c}\psi_{b}^{\dagger}]}

which may be written in the form of a Lax equation

\eqn\FLaxExplicit
{\partial_{+}\partial_{-}^{2}\sigma ={ig^{2}\over\sqrt{2}}
[\partial_{-}^{2}\sigma ,\sigma].}

This last equation is nothing but the conservation of the
probability 2-current. However, due to the fact that the
trace of the $-$ component of the probability current
vanishes, the trace of any power of the $+$ component of the probability
current is conserved (with respect to the "time" $x^{+}$).
Since this quantity is local, it furnishes an infinite number
of conserved quantities, indexed by the continous label $x^{-}$.
Of course, this conservation law arises from the usual global
$U(1)$ symmetry. What is suprising however, is that in this
case, it is not the total charge (i.e. the integrated charge
density) that is conserved, but rather the charge density
itself. Note the close
similarity between \FLaxExplicit\ and the dynamical equation
for the Heisenberg spin chain

\eqn\HSChain
{\partial_{+}S=ig^{2}[\partial_{-}^{2}S,S],}

where $S$ may be expanded in the basis of generators of $SU(2)$.
A close connection between the Heisenberg spin chain and $SU(2)$
chiral $QCD_{2}$ will emerge when we perform a systematic inverse
scattering treatment in a later section.

Consider next the time dependance of $\partial_{-}\sigma_{ab}(x)$.
From the definition \Sigm\

\eqn\DerivSigm
{\eqalign{\partial_{+}\partial_{-}\sigma_{ab}(x)
&=\partial_{+}\int dy^{-} \epsilon (x^{-}-y^{-})
\psi_{a}(y^{-})\psi_{b}^{\dagger}(y^{-})
\cr\noalign{\vskip 0.2truecm}
&={ig^{2}\over\sqrt{2}}
\int dy^{-}\epsilon (x^{-}-y^{-})[\partial_{-}^{2}\sigma,\sigma]_{ab}.}}

Now, noticing that

\eqn\OutDer
{\partial_{-}[\partial_{-}\sigma(x),\sigma(x)]
=[\partial^{2}_{-}\sigma(x), \sigma(x)]}

we find, after integrating by parts and dropping the boundary term

\eqn\DerivSigmTwo
{\partial_{+}\partial_{-}\sigma_{ab}(x)={ig^{2}\over\sqrt{2}}
[\partial_{-}\sigma(x), \sigma(x)]_{ab}}

which is again of the Lax form. This Lax pair again gives an
infinite number of conserved quatities contained in the
conservation law

\eqn\ChargesTwo
{\partial_{+} Tr \Big(\partial_{-}\sigma(x)\Big)^{n}=0}

The conserved charges discussed above are all gauge invariant 
quantities. Note that the equations 
\DerivSigmTwo\ and \FLaxExplicit\ are invariant under
diffeomorphisms of the form $x^{-}\to f(x^{-})$ \rUs\ . The action however
is not invariant under this symmetry. 
This diffeomorphism invariance will be used in the next
section to provide a geometric interpretation of the problem.

Finally we comment that it is also a simple task to show that 

\eqn\LastCharge
{\partial_{+} Tr \Big(\sigma^{n}(x)\Big)=0}

which again represents an infinite number of conserved quantities. The 
conserved quantities \ChargesTwo\ and \LastCharge\ are both non-local
functionals of the original field variables. For a discussion on 
non-local conservation laws for the Heisenberg magnet, see \rJevicki\ .

\subsec{The $r$ matrix.}

The existence of an infinite sequence of conservation laws does not
automatically guarantee integrability of the classical dynamics. One
must prove in addition, that these conserved quantities Poisson
commute. A very elegant way in which this is demonstrated is due to
Faddeev \rFaddeev\ . The idea is to introduce the $r$ matrix, defined by the
fundamental Poisson relation (FPR)

\eqn\FPR
{\{ U(x,\lambda)\matrix{\otimes\cr ,}U(y,\mu )\}
=\big[ r(\lambda-\mu ),U(x,\lambda)\otimes I+I\otimes
U(y,\lambda )\big]\delta (x-y),}

where we have introduced the notation

\eqn\NotFFPR
{\{ A\matrix{\otimes\cr ,}B\}_{ik|lm}=\{A_{il},B_{km}\}.}

Usually the FPR is used to prove that certain conserved
monodromies Poisson commute. The treatment for chiral $QCD_{2}$
is much simpler. Considering
$U(x,\lambda )={i\over 2\lambda}\partial_{-}^{2}\sigma (x)$, and
taking traces of the above equation immediatly implies that

\eqn\PoissonCommute
{\{ Tr(\partial_{-}^{2}\sigma)(x),Tr(\partial_{-}^{2}\sigma)(y)\} =0.}

Thus, if we can find an $r$ matrix such that \FPR\ holds, then
we have demonstrated that our classically conserved quantities Poisson
commute. For the case of the gauge group $SU(2)$, it is not difficult
to verify that $\partial_{-}^{2}\sigma(x)$ obeys exactly the same
Poisson bracket relations as the Heisenberg spin chain variables
$\sigma^{a}S^{a}(x)$. In addition, the $r$ matrix for the Heisenberg
spin chain, where $U(x,\lambda)={i\over 2\lambda}\sigma^{a}S^{a}(x)$
is known. Thus, the $r$ matrix defined by \FPR\ is nothing but the $r$
matrix of the Heisenberg spin chain! In a natural basis, the $r$ matrix
is given by  $r=P/(\lambda -\mu)$ \rFaddeev\ where

\eqn\rmatrix
{P=\left[\matrix{1\quad 0\quad 0\quad 0\cr
0\quad 0\quad 1\quad 0\cr
0\quad 1\quad 0\quad 0\cr
0\quad 0\quad 0\quad 1\cr}\right].}

\subsec{Zero Curvature Condition.}

We begin by noting that the integrability condition of the linear equations

\eqn\LinOne
{\partial_{-}\sigma_{ab}(x)\phi_{b} (x)=\alpha\phi_{a} (x)}

\eqn\LinTwo
{\partial_{+}\phi_{a}=-{ig^{2}\over\sqrt{2}}\sigma_{ab}(x)\phi_{b}(x)}

reproduces \DerivSigmTwo\ . Now, shifting $x^{-}\to x^{-}+\eta$ 
and using first
order perturbation theory to compute $\delta\phi=\eta\partial_{-}\phi$
we are led to

\eqn\ZerCurTwo
{\partial_{-}\phi_{a} = -{ig^{2}\over\sqrt{2}}x^{+}\partial_{-}\sigma_{ab}(x)
\phi_{b}.}

Clearly, the pair of equations \ZerCurTwo\ and \LinTwo\ , which constitute
the linear problem, also have the diffeomorphism symmetry. The integrability
condition of \ZerCurTwo\ and \LinTwo\ now provide a zero curvature
representation

\eqn\ZerCurvOne
{[\partial_{+}+{ig^{2}\over\sqrt{2}}\sigma ,\partial_{-}+
{ig^{2}\over \sqrt{2}}x^{+}\partial_{-}\sigma ]=0}

of \DerivSigmTwo\ . This zero curvature representation is well suited to a
systematic inverse scattering treatment, which is provided in section 4.

\newsec{The case of SU(2): Some Preliminary Comments.}

In this section we specialize our discussion to the case $N=2$.

\subsec{The Geometry of the Lax Pair.}

Expanding the Lax equation \DerivSigmTwo\ in the basis provided
by the Pauli matrices ($\sigma^{i}$) as

\eqn\ExpSigm
{\sigma_{ab}(x)={1\over 2}f^{i}(x)\sigma^{i}_{ab}}

yields

\eqn\fdyn
{\partial_{+}\partial_{-}f^{i}=-{g^{2}\over
\sqrt{2}}\epsilon^{ijk}\partial_{-}f^{j}
f^{k}}

After using the diffeomorphism symmetry to change to the new
variable $s(x^{-})$, such that
$\partial_{s}f^{i}\partial_{s}f^{i}=1$, we are able to interpret
$s$ as the arclength of the curve traced out by the three
co-ordinates $(f^{1},f^{2},f^{3})$ at any given time $x^{+}$.
Given the arclength and position of the curve it is a simple matter
to compute the explicit expressions

\eqn\Nor
{(e_{1})^{i}=\partial_{s}f^{i}}

\eqn\BiNor
{(e_{2})^{i}={1\over\kappa}\partial_{s}^{2}f^{i}}

\eqn\OscNor
{(e_{3})^{i}={1\over\kappa}\epsilon^{ijk}\partial_{s}f^{j}
\partial_{s}^{2}f^{k},}

for the normal, binormal and osculating normal and

\eqn\Cur
{\kappa =\sqrt{\partial_{s}^{2}f^{i}\partial_{s}^{2}f^{i}}}

\eqn\Tor
{\tau ={1\over\kappa^{2}}\epsilon^{ijk}\partial_{s}f^{i}\partial_{s}^{2}f^{j}
\partial_{s}^{3}f^{k},}

for the curvature and the torsion. The dynamics of the three unit
normals $e_{i}$ is easily found, upon using \fdyn\

\eqn\NorDyn
{\partial_{+}(e_{n})_{i}=-\sqrt{2}g^{2}\epsilon^{ijk}f^{j}(e_{n})_{k}}

In addition to this, the propagation of the curve in arclength is
described by the Frenet-Serret equations which read

\eqn\SFOne
{\partial_{s}e_{1}=\kappa e_{2}}

\eqn\SFTwo
{\partial_{s}e_{2}=\tau e_{3}-\kappa e_{1}}

\eqn\SFThree
{\partial_{s}e_{3}=-\tau e_{2}.}

\subsec{A Second Zero Curvature Condition.}

Now, transforming to Darboux co-ordinates

\eqn\Dar
{z_{l}={(e_{2})_{l}+i(e_{3})_{l} \over 1-(e_{1})_{l}}\quad
(e_{1})^{2}_{l}+(e_{2})^{2}_{l}+(e_{3})^{2}_{l}=1}

we are led to the pair of Riccati equations

\eqn\Rone
{\partial_{-}z_{l}=-i\tau z_{l}+{\kappa\over 2}(z_{l}^{2}+1)}

\eqn\Rtwo
{\partial_{+}z_{l}=-iaz_{l}+{1\over 2}(b+ic)z_{l}^{2}+{1\over 2}
(b-ic),}

where

\eqn\adef
{a=-if^{i}\partial_{s}f^{i},}

\eqn\bdef
{b={1\over \kappa\tau}\left[ f^{i}\partial_{s}^{3}f^{i}-{1\over\kappa^{2}}
\partial_{s}^{2}f^{j}\partial_{s}^{3}f^{j}f^{i}\partial_{s}^{2}f^{i}
+\kappa^{2}f^{i}\partial_{s}f^{i}\right]}

and

\eqn\cdef
{c={1\over\kappa}f^{i}\partial_{s}^{2}f^{i}.}

The Riccati equations derived above, and their relation to the Frenet frame
will prove to be vital for the construction of auto-B\"acklund
transformations. It is simple to verify that this is simply the coset space
representation $z={v_{1}\over v_{2}}$ of the following pair of linear
eigenvalue problems

\eqn\linearOne
{\left[
\matrix{
\partial_{s}\beta_{1} \cr \partial_{s}\beta_{2} \cr}
\right]=\left[
\matrix{
{i\over 2}\tau &  -{\kappa\over 2}\cr 
{\kappa\over 2} & {-i\over 2}\tau}
\right]\left[
\matrix{
\beta_{1} \cr\beta_{2} }\right]=U_{ij}\beta_{j}}

\eqn\linearTwo
{\left[
\matrix{
\partial_{+}\beta_{1} \cr \partial_{+}\beta_{2} \cr}
\right]=\left[
\matrix{
{i\over 2}a & -{1\over 2}(c-ib) \cr {1\over 2}(c-ib)
 & {-i\over 2}a}
\right]\left[
\matrix{
\beta_{1} \cr\beta_{2} }\right]=V_{ij}\beta_{j}.}

The integrability condition for this pair of equations
implies\foot{These restrictions on the curvature and the
torsion can also be obtained directly as the integrability
condition for the equations of motion \fdyn\ and
the Frenet-Serret equations \rUs\ .}

\eqn\Condition
{\partial_{+}\tau=\sqrt{2}g^{2} \quad \partial_{+}\kappa =0.}

This last result is remarkable and unexpected. It implies
that {\it any} curve may be mapped into a solution of $QCD_{2}$.

The connection between soliton solutions and curves has been
discussed by a number of authors \rHas\ . In particular,
Lamb \rLamb\ shows how the single soliton solution of the
Sine-Gordon model can be related to a curve of curvature
$\kappa =a sech(as)$ and torsion $\tau =$ constant. This
curve, upon identifying
$\tau =\sqrt{2}g^{2}x^{+}$, provides a soliton solution
of the equation \fdyn\ . Explicitely, we have

\eqn\CurveSoln
{\eqalign{f_{1}&={2a \over 2g^{4}x^{+2}+a^{2}}sech(as)
sin(\sqrt{2}g^{2}sx^{+})
\cr\noalign{\vskip 0.2truecm}
f_{2}&= {2a \over 2g^{4}x^{+2}+a^{2}}sech(as)
cos(\sqrt{2}g^{2}sx^{+})
\cr\noalign{\vskip 0.2truecm}
f_{3}&= s-{2a\over 2g^{4}x^{+2}+a^{2}}tanh(as).}}

It is a straightforward excercise to explicitly verify that these
$f_{i}$ do indeed satisfy the equation of motion \fdyn\ . We also
state the solution with constant curvature $\kappa$ and torsion
$\tau={g^{2}\over\sqrt{2}}x^{+}$

\eqn\CurveSln
{ \eqalign{f_{1}&={2\kappa \over g^{4}x^{+2}+2\kappa^{2}}
sin(\sqrt{\kappa^{2}+{g^{4}\over 2}x^{+2}}s)
\cr\noalign{\vskip 0.2truecm}
f_{2}&= {2\kappa \over g^{4}x^{+2}+2\kappa^{2}}
cos(\sqrt{\kappa^{2}+{g^{4}\over 2}x^{+2}}s)
\cr\noalign{\vskip 0.2truecm}
f_{3}&= -{g^{2}x^{+}s\over\sqrt{g^{4}x^{+2}+2\kappa^{2}}}.}}

\subsec{An Auto-B\"acklund Transformation.}

In this section, we will concentrate our attention on the class of
constant torsion curves. Of course, this is only a small subset of all
possible solutions to our equations of motion, but as we shall see, it
includes a number of interesting solutions.

The geometrical interpretation of a B\"acklund transformation is as a
transformation that takes a given pseudospherical surface\foot{i.e. a
surface with constant negative Gauss curvature.} to a new pseudospherical
surface. Moreover, a B\"acklund transformation takes asymptotic lines to
asymptotic lines. The fact that asymptotic lines on a pseudospherical
surface have constant torsion, is the first hint that it is possible to
restrict the B\"acklund transformation to get a transformation that
carries constant torsion curves to constant torsion curves. That this
is indeed possible is the result of a theorem
due to Calini and Ivey\rCalini\ :
Assume that the osculating normal($e_{3}$), binormal ($e_{2}$) and tangent
($e_{1}$) vectors to a constant torsion ($\tau$) curve, together with
the curves curvature ($\kappa$) are known. The curve with Frenet frame
given by

\eqn\BTFrameOne
{e_{1}'=e_{1}+(1-cos\theta )sin\beta (cos\beta e_{2}-sin\beta e_{1})+
sin\theta sin\beta e_{3},}

\eqn\BTFrameTwo
{e_{2}'=e_{2}-(1-cos\theta )cos\beta (cos\beta e_{2}-sin\beta e_{1})-
sin\theta cos\beta e_{3},}

\eqn\BTFrameThr
{e_{3}'=cos\theta e_{3}+sin\theta (cos\beta e_{2}-sin\beta e_{1}}

where $\beta$ solves the equation

\eqn\EFBeta
{{d\beta\over ds}=Csin\beta -\kappa,}

and $tan({\theta\over 2})={C\over\tau}$, has torion $\tau'=\tau$, and
curvature

\eqn\NCur
{\kappa'= \kappa-2Csin\beta.}

From the results of the previous section, this clearly implies that
the above transformation constitutes an auto-B\"acklund transformation
for $QCD_{2}$.

Let us illustrate this B\"acklund transformation with an example. First,
lets find the B\"acklund transformation of the "trivial" solution of
\SFOne\ - \SFThree\ , i.e. the solution
with $\kappa=0$ and $\tau=\sqrt{2}g^{2}x^{+}$. A suitable Frenet
frame for this curve is given by

\eqn\SFFfzc
{e_{1}=\left[\matrix{0\cr 0\cr -1}\right],\quad
e_{2}=\left[\matrix{sin(\sqrt{2}g^{2}x^{+}s)
\cr cos(\sqrt{2}g^{2}x^{+}s)\cr 0}\right],\quad
e_{3}=\left[\matrix{cos(\sqrt{2}g^{2}x^{+}s)\cr
-sin(\sqrt{2}g^{2}x^{+}s)\cr 0}
\right].}

For $\kappa=0$,\EFBeta\ has the solution

\eqn\FRCr
{\beta =-2arctan(e^{Cs+\alpha})}

which corresponds to a curve with curvature
$\kappa'=-2Csin\beta=2asech(Cs+\alpha )$. The tangent
to this B\"acklund transformed curve follows directly
from \BTFrameOne\

\eqn\ttsih
{e_{1}=\left[\matrix{
{-2a^{2}\over a^{2}+2g^{4}x^{+2}}
sech(as)tanh(as)sin(\sqrt{2}g^{2}x^{+}s)
-{2\sqrt{2}ag^{2}x^{+}\over a^{2}+2g^{4}x^{+2}}
cos(\sqrt{2}g^{2}x^{+}s)\cr
{-2a^{2}\over a^{2}+2g^{4}x^{+2}}
sech(as)tanh(as)cos(\sqrt{2}g^{2}x^{+}s)
+{2\sqrt{2}ag^{2}x^{+}\over a^{2}+2g^{4}x^{+2}}
sin(\sqrt{2}g^{2}x^{+}s)\cr
1-{2a\over a^{2}+2g^{4}x^{+2}}sech^{2}(as)
}\right].}

This is nothing but the soliton solution \CurveSoln\
we found earlier.

It is clear that in order to implement the B\"acklund transformation
we need to solve the associated equation \EFBeta\ , which is difficult for
non trivial curvatures. In the remainder of this section, we describe
how to solve \EFBeta\ for any curve whose Frenet frame is explicitly
known and can be continued analytically as a function of $\tau$, with fixed
curvature.

First, setting $y=tan{\beta\over 2}$, \EFBeta\ becomes

\eqn\NREqtion
{{dy\over ds}=ay-{\kappa\over 2}(1+y^{2}).}

Now, if we analytically continue the (real) torsion to $\tau =-ia$, then we
find that we can identify $y=-z_{l}$ where $z_{l}$ is a Darboux coordinate
\Dar\ satisfying the Ricatti equation \Rone\ . This provides a direct way of
obtaining $\beta$.

Lets illustrate this last point with an example. From the Frenet frame
\SFFfzc\ we construct the Darboux coordinate $z_{2}=cos(\tau x)+isin(\tau x).$
Evaluating this at $\tau =-ia$ yeilds $y=-e^{ax}$, which implies
$\beta=-2arctan(e^{ax})$ in agreement with \FRCr\ .

\subsec{Fermion Number.}

Our classical solutions should correspond to solutions with a well defined
fermion number. Thus, we should be able to identify an integer valued charge
associated with each of the above classical solutions. In this section we
show that this fermion number arises in a very natural way.

First, note that from (2.3), it is clear that the quark field $\psi_{a}$ is
an eigenvector of $\partial_{-}^{2}\sigma$ with eigenvalue equal to the
fermion number. Since $\partial_{-}^{2}\sigma$ is a two dimensional traceless
Hermittian matrix, it has two real eigenvalues of opposite sign. The positive
eigenvalue corresponds to the fermion number density. Thus, we may write

\eqn\TFNdensity
{\psi_{a}^{\dagger}\psi_{a}=\sqrt{{1\over 2}Tr(\partial_{-}^{2}\sigma)^{2}}.}

In the curve language, this is simply the curvature. Thus, the total fermion
number should be proportional to the integral over arclength of the 
curvature of the solution. 

Now, consider the set of solutions that arise as B\"acklund transformations
of the solution with $\kappa=0$ . These solutions have curvature

\eqn\Curvey
{\kappa=-2Csin(\beta ),}

where

\eqn\EFllBeta
{{d\beta\over ds}=Csin(\beta ).}

To obtain a soliton solution, we seek solutions which have the property
${d\beta\over ds}\to 0$ as $s\to\pm\infty$ . From the above equation, this 
tells us that $\beta\to n\pi$ as $s\to\pm\infty$ for any integer $n$ . Thus,
the total fermion number is proportional to

\eqn\TFNhere
{\int ds\kappa(s)=-\int ds 2Csin(\beta )=-2\int ds {d\beta\over ds}=
-2\beta |^{\infty}_{-\infty}=-2(n_{1}-n_{2})\pi.}

Thus, the fermion number for these solutions are an integer times $2\pi $ .
Now, consider the solution obtained by B\"acklund transformation of the
above solution. In this case the curvature is given in (3.27) where the
dynamics of $\beta$ is now (3.26). The boundary conditions for $\beta$ are 
unchanged, so that the above argument again leads to the conclusion that
the total fermion number is an integer multiplied by $2\pi$. Indeed, any
solution which is obtained from a sequence of B\"acklund transformations
of the $\kappa=0$ solution, have a fermion number equal to $2\pi$
multiplied by an integer. The above arguments can be generalized to
solutions obtained using the auto-B\"acklund transformation on a curve
with an arbitrary initial curvature. In this case, the fermion number
of the solutions obtained through the B\"acklund transformation will
differ from the fermion number of the original curve by $2\pi n$ where
$n$ is an arbitrary integer. There is however no restriction on the
fermion number of the initial curve.
Note that the results discussed in this section depend on the fact that
$x^{-}$ is identified with the arclength along the curve.

\subsec{Closed Curves.}

In this section, we consider the following differential-geometric problem: If
a curve in $R^{3}$ is closed, then the curvature and torsion are necessarily
periodic. However, given a periodic curvature and a periodic torsion, we do
not in general obtain a closed in curve. In this section, we will show that
the extra constraints that must be placed on the curvature and the torsion
so as to obtain a closed curve,
are naturally expressed in terms of the spectral problem for $QCD_{2}$. More
precisely, the requirement that a curve is periodic can be expressed as a
constraint on the spectral curve of classical chiral $SU(2)$ $QCD_{2}$. It
is interesting to study this curve, since spectral curves associated with
other $(1+1)$ dimensional models have been used to derive the Seiberg-Witten
differential (and therefore also the exact mass-charge relationship of the
BPS saturated states appearing in the low energy dynamics \rMarsh\ ). 
The observations made in this section are based on the recent 
proof given in \rcc\ where
Riemann surfaces corresponding to periodic curves are identified.

As a start, consider the zero curvature condition given in 
\linearOne\ and \linearTwo\ . Since the
zero curvature condition is invariant under gauge transformations, we can
equivalently consider the spectral problem\foot{In what follows, we have
normalized the length of the closed curve (given by traversing the curve
once) to $2\pi$.}

\eqn\GTSpect
{\eqalign{U\to \tilde{U}&=g^{-1}Ug-g^{-1}\partial_{s}g,\cr
V\to \tilde{V}&=g^{-1}Vg-g^{-1}\partial_{x^{+}}g,}}

where the connection $U,V$ are defined in 
\linearOne\ and \linearTwo\ , and

\eqn\NewConnect
{g=\left[
\matrix{exp({i\over 2}\int^{x^{-}}_{0}dx'\tau (x'))\quad
0\cr 0\quad exp(-{i\over 2}\int^{x^{-}}_{0}dx'\tau (x'))}\right].}

Although this gauge transformation changes the connections $U$ and $V,$ it
leaves the integrability condition unchanged. Concentrate on the problem

\eqn\RelProb
{L(\lambda )F(s,\lambda)=0=\Big({d\over ds}-U\Big)F(s,\lambda ),
\quad L(\lambda )={d\over ds}-\left[
\matrix{-{1\over 2}i\lambda\quad {1\over 2}iq(s)\cr
{1\over 2}i\bar{q}(s)\quad {1\over 2}i\lambda}\right].}

The crucial elements of the above problem are that the magnitude of $q(s)$ is
related to the curvature of the curve, and that the derivative with respect
to arclength of the phase of $q(s)$ is related to the torsion. It is this
property of $QCD_{2}$ that makes it possible to consider the closed curve
problem. The link just described, between a curve and a function $q(s)$ is
known as the Hasimoto map\rHas\ . In what follows, we will need to make use of 
the
shift operator, defined by the action

\eqn\ShiftOp
{f(s)\to f(s+2\pi ).}

Here, since we are looking for a closed curve, both the torsion and curvature
are periodic. It thus makes to sense to think that our quarks will be moving
(at fixed $x^{+}$) in a periodic potential. In analogy to the treatment of
waves moving through a periodic potential (lattice) we define Bloch functions
as the functions which are eigenfunctions of both the shift operator and
$L(\lambda )$

\eqn\BlochFuncs
{\eqalign{
\psi^{(1)}&=\left[
\matrix{\psi_{1}^{(1)}(s,\lambda )\cr \psi_{2}^{(1)}(s,\lambda )}\right],
\quad \psi^{(2)}=\left[
\matrix{\psi_{1}^{(2)}(s,\lambda )\cr \psi_{2}^{(2)}(s,\lambda )}
\right],\cr
L(\lambda )\psi^{(1)}(s,\lambda )&=0,\quad \psi^{(1)}(s+2\pi,\lambda )=
\omega^{(1)}(\lambda )\psi^{(1)}(s,\lambda ),\cr
L(\lambda )\psi^{(2)}(s,\lambda )&=0,\quad \psi^{(2)}(s+2\pi,\lambda )=
\omega^{(2)}(\lambda )\psi^{(2)}(s,\lambda ).}}

Since $U(s,\lambda )$ is traceless, the normalization of the Bloch functions
does not change with arclength $s$, so that we must have

\eqn\NormCond
{\omega^{(1)}(\lambda )\omega^{(1)*}(\lambda )
=\omega^{(1)}(\lambda )\omega^{(2)*}(\lambda)=1.}

A point in the comple plane $\lambda$ belongs to the spectrum of
$L(\lambda)$ if and only if $|\omega^{(1)}(\lambda )|=1$. Thus,
we have obtained
our first characetrization of our spectral curve. Now, we will use the
concept of a transition matrix, which is introduced in the next section. The
reader is asked to consult that section for details.\foot{We trust that this
will not confuse the reader.} For a generic complex $\lambda$, the transition
matrix $T(\lambda )$ will have two eigenvalues $\omega^{(1)}(\lambda)$ and
$\omega^{(2)}(\lambda)$ and a pair of corresponding Bloch functions. In terms
of these two functions, we introduce a hyperelliptic Riemann surface\foot{ A
Riemann surface is called hyperelliptic if it is a two sheeted ramified
covering of the Riemann sphere.} $Y$ such
that $\omega^{(1)}(\lambda)=\omega(\mu_{1})$ and
$\omega^{(2)}(\lambda)=\omega(\mu_{2})$ where $\mu_{1}$ and $\mu_{2}$ are the
pre-images of the point $\lambda$ under the projection $Y\to C$. Note that
any hyperelliptic Riemann surface obviously has a natural holomorphic
involution which amounts to transposing the two sheets. The function

\eqn\QuasMom
{p(\mu )={1\over 2\pi i}ln\omega (\mu )}

is called the quasimomentum function. In what follows, the quasimomentum
differential

\eqn\QuasMomDif
{dp(\mu )={d p(\mu)\over d\mu} d \mu}

will play an important role.

The structure of the Riemann surface $Y$ (called the Bloch variety) has been
studied in detail in \rY\ . We will need the following definitions: A complex
point $\lambda$ is called regular if
$\omega^{(1)}(\lambda)\ne\omega^{(2)}(\lambda),$ and irregular if
$\omega^{(1)}(\lambda)=\omega^{(2)}(\lambda).$ There are three types of
irregular points (1) Branch points, (2) Non-removable double points and
(3) Removable double points. An irregular point $\lambda_{0}$ is called a
branch point if in going around this point we move from one sheet of $Y$ to
another (i.e. the monodromy around this point is non-trivial). If the
monodromy around an irregular point is trivial, then this point is a double
point. If the Bloch functions are equal at the double point, then it is
non-removable; if the Bloch functions are not equal at the double point, then
the double point is removable.

One last piece of notation is needed now. Introduce the new $SU(2)$ valued
frame $\sigma_{i}\hat{E}^{i}_{n}$, related to the Frenet frame by

\eqn\NewFrme
{\eqalign{\hat{E}_{1}&=e_{1},\cr
\hat{E}_{2}&=cos(\theta )e_{2}-sin(\theta )e_{3},\cr
\hat{E}_{3}&=sin(\theta )e_{2}-cos(\theta )e_{3},\cr
\theta&=\int^{s}\kappa (s')ds'.}}

The $SU(2)$ valued coordinate of the curve is now given by

\eqn\NewCurveCoord
{\hat{\Gamma}(s)=\hat{\Gamma}(0)+\int_{0}^{s}ds'\hat{E}_{1}(s').}

We are now ready to state the theorem that provides the restriction on the
spectral curve: Let $q(s)$ be a complex valued smooth periodic function of
one variable $s$, $q(s)\ne 0$ for all $s$, $q(s+2\pi )=q(s)$, $\Lambda_{0}$
an arbitrary real number and $\hat{\Gamma}(s)$ the corresponding curve
constructed from the curvature and torsion specified in $q(s)$. Then, (1) the
matrix $\hat{E}_{1}(s)$ is periodic with period $2\pi$, if and only if
$\Lambda_{0}$ is a double point of the Bloch variety $Y$ and (2) the function
$\hat{\Gamma}(s)$ is periodic with period $2\pi$ if and only if $\Lambda_{0}$
is a double point of $Y$ and $dp(\mu_{1})=0=dp(\mu_{2})$ where $\mu_{1}$ and
$\mu_{2}$ are the preimages of $\Lambda_{0}$ under the projection
$Y\to C$. We are content with the statement of the theorem, and refer the
reader to \rcc\ for a proof.

To summarize, we have managed to show that the spectral problem in $QCD_{2}$
can be used to define Bloch functions for periodic potentials. The
requirement that these periodic potentials actually correspond to closed
curves is then easily expressed as a condition on the quasimomentum
differential.

\subsec{A Nonlinear Superposition Principle.}

The usual approach to massive $QCD_{2}$ is to begin by performing a non-Abelian
bosonization. Then, classical solutions of the bosonic action are found and
a semi classical quantization may be carried out. The static classical 
solutions
of these bosonic equations satisfy the Sine-Gordon equation. This allows the
construction of multi-soliton solutions using the well known nonlinear
superposition associated with the Sine-Gordon equation \rRo\ .
In this section, we
show that the nonlinear superposition principle also plays a role for the 
chiral
theory. This explicitly 
demonstrates the existence of multi soliton solutions,
which is another signal of the theories classical integrability.

Begin by considering the 
spatial part of the B\"acklund transformation for the
Sine-Gordon equation

\eqn\SinGordBT
{{\partial (\tilde{\varphi}-\varphi)\over\partial s}=C sin(\tilde{\varphi}
+\varphi ),}

where $\varphi$ is the one soliton solution and hence solves

\eqn\SGOneSoliton
{{\partial\varphi\over\partial s}=sin\varphi.}

From this last equation, 
it is clear that the single soliton solution corresponds
to a curve with curvature $-2sin\varphi$. Now, noting that

\eqn\SGTwoSoliton
{{\partial(\varphi+\tilde{\varphi})\over\partial s}=
Csin(\varphi+\tilde{\varphi})+2{\partial\varphi\over\partial s}
=Csin(\varphi+\tilde{\varphi})+2Csin(\varphi)}

we see that we may identify $\beta=\varphi+\tilde{\varphi}$ to obtain a
B\"acklund transformation to a curve with curvature
$-2sin\varphi-2sin\beta$. Now, lets compute the three soliton solution
$\bar{\varphi}$ from the two soliton solution $\tilde{\varphi}$

\eqn\SGThreeSoliton
{{\partial (\bar{\varphi}-\tilde{\varphi})\over\partial s}=
Csin(\bar{\varphi}+\tilde{\varphi}).}

Rewriting this equation as

\eqn\AnotherBT
{{\partial (\bar{\varphi}+\tilde{\varphi})\over\partial s}=
Csin(\bar{\varphi}
+\tilde{\varphi})+2{\partial\tilde{\varphi}\over\partial s}
=Csin(\bar{\varphi}+\tilde{\varphi})+2(Csin(\varphi+\tilde{\varphi})
+sin(\varphi)),}

so that identifying $\beta=\bar{\varphi}+\tilde{\varphi}$ provides a
B\"acklund transformation to a curve of curvature
$-2sin(\varphi+\bar{\varphi}+\tilde{\varphi})
-2sin(\varphi+\tilde{\varphi})-2sin(\varphi)$.
By now, the generalization to the $n$ soliton solution is obvious.


From the nonlinear superposition for the Sine-Gordon equation, we obtain
the algebraic formula

\eqn\nonlinsuppos
{\beta=2arctan\Big[{c_{1}+c_{2}\over c_{1}-c_{2}} tan\Big(
{\phi^{1}_{n}-\phi^{2}_{n}\over 2}\Big)\Big]+\phi_{n}^{1}+\phi_{n-1},}

for the B\"acklund transformation 
contructed from the $n+1$ soliton solution
of the Sine-Gordon equation. In this last formula, $\phi_{n}^{1}$ is the
$n$ soliton solution (of the Sine-Gordon equation)constructed from the
$n-1$ soliton solution $\phi_{n-1}$ (of the Sine-Gordon equation)
with parameter $c_{1}$, $\phi_{n}^{2}$ is the $n$ soliton solution
(of the Sine-Gordon equation) constructed
from $\phi_{n-1}$ with parameter $c_{2}$. Thus, the nonlinear superposition
principle for the Sine-Gordon equation can be used to construct the multi
soliton solutions of chiral $QCD_{2}$.

\newsec{Dynamical Analysis using the Inverse Scattering Method.}

In this section, we present a systematic inverse scattering treatment of the
classical equations of motion for the gauge group $SU(2)$. We reproduce the
single soliton solution described in the previous section, clearly
demonstrating the validity of the analysis. However, in contrast to the
previous section, all results in this section are easily generalized to
$SU(N)$.

\subsec{The direct scattering problem.}

In this section, we study the linear problem (2.22) derived in section 2.3.
The eigenvalue problem of the Lax operator reads

\eqn\LinProb
{\partial_{-}\phi_{a}=\lambda \partial_{-}\sigma_{ab}\phi_{b}. }

From the time dependance of $\partial_{-}\sigma$, \DerivSigmTwo\ ,
it is clear that requiring that $\partial_{+}\partial_{-}\sigma\to 0$
as $|x^{-}|\to\infty$ implies that the commutator of $\sigma$ and
$\partial_{-}\sigma$ must vanish as $|x^{-}|\to\infty$. There are
two ways in which this commutator may vanish. First, the field
$\partial_{-}\sigma$ itself vanishes. Since we are working in terms
of the arclength variable $s$, we have

\eqn\BC
{(e_{1})_{i}(e_{1})_{i}={1\over 2}Tr(\partial_{-}\sigma^{2})=1,}

so that $\partial_{-}\sigma\ne 0$ as $|x^{-}|\to\infty$. The second
possibility is that $\sigma$ becomes proportional to a single
Pauli matrix. Without loss of generality (thanks to global color
rotation invariance), we may take $\sigma$ to be proportional to
$\sigma_{z}$. The requirement \BC\ forces
$\partial_{-}\sigma\to\sigma_{z}$ as $|x^{-}|\to\infty$,
which fixes the relevant boundary condition for \LinProb\ .
This linear problem has been studied by Takhtajan \rTak\ (see also
Fogedby \rFog\ ) in connection with the classical Heisenberg spin chain.
In this section we will review the results of his work relevant
for the present analysis.

The two Jost solutions $F(x,\lambda )$ and $G(x,\lambda )$ defined
by the boundary conditions

\eqn\BCOne
{F(x,\lambda )\to exp(-i\lambda\sigma_{3}x)}

as $x\to\infty$ and

\eqn\BCTwo
{G(x,\lambda )\to exp(-i\lambda\sigma_{3}x)} 

as $x\to -\infty$, have the integral representations

\eqn\IntOne
{F(x,\lambda )=exp(-i\lambda\sigma_{3}x)+\lambda\int^{\infty}_{x}K(x,y)
exp(-i\lambda\sigma_{3}y)dy}

\eqn\IntTwo
{G(x,\lambda )=exp(-i\lambda\sigma_{3}x)+\lambda\int^{x}_{-\infty}N(x,y)
exp(-i\lambda\sigma_{3}y)dy.}

The kernels $K(x,y)$ and $N(x,y)$ are independant of the eigenvalue
$\lambda$ and solve the Goursat problem

\eqn\GPOne
{{\partial K(x,y)\over\partial x}\sigma_{3}+\partial_{-}\sigma (x)
{\partial K(x,y)\over\partial y}=0 \quad (x\le y),}

\eqn\GPTwo
{{\partial N(x,y)\over\partial x}\sigma_{3}+\partial_{-}\sigma (x)
{\partial N(x,y)\over\partial y}=0 \quad (x\ge y),}

with the boundary conditions

\eqn\GBCOne
{\partial_{-}\sigma (x)-\sigma_{3}+iK(x,x)-i\partial_{-}\sigma (x)K(x,x)
\sigma_{3} =0,}

\eqn\GBCTwo
{\partial_{-}\sigma (x)-\sigma_{3}+iN(x,x)-i\partial_{-}\sigma (x)N(x,x)
\sigma_{3} =0.}

These last two boundary conditions imply that 

\eqn\GPCone
{\partial_{-}\sigma (x)=(iK(x,x)-\sigma_{3})\sigma_{3}(iK(x,x)-
\sigma_{3})^{-1},}

and

\eqn\GPCtwo
{\partial_{-}\sigma (x)=(iN(x,x)-\sigma_{3})\sigma_{3}(iN(x,x)-
\sigma_{3})^{-1}.}

The linear system \LinProb\ has the important property that if
$\Psi_{1}$ and $\Psi_{2}$ are solutions, then $\Psi_{2}=\Psi_{1}A$,
with $A$ a constant.\foot{This property is easily proved by
differentiating $\Psi_{2}^{-1}\Psi_{1}$ with respect to $x^{-}$,
and using \LinProb\ .} This allows us to write

\eqn\ScatMat
{G(x,\lambda )=F(x,\lambda )T(\lambda ),}

where $T(\lambda )$, is the transition matrix. If $\Psi$ is a
solution of \LinProb\ , then writing $det(\Psi )$ as
$exp(Trln\Psi )$ we find, upon using $Tr(\partial_{-}\sigma) =0$,
that $det(\Psi )$ is independant of $x^{-}$. Also, since the Pauli
matrices have the property that
$\sigma_{i}^{*}=-\sigma_{2}\sigma_{i}\sigma_{2}$,
it is clear that if $\Psi$ is a solution of \LinProb\ for a
spectral parameter $\lambda$, then $\sigma_{2}\Psi^{*}\sigma_{2}$ is a
solution for $\lambda^{*}$. Using these properties of the
solutions to \LinProb\ , it is not difficult to show that the
transition matrix has the form

\eqn\TransMat  
{T(\lambda )=\left[
\matrix{
a(\lambda ) &  -(b(\lambda^{*})^{*} \cr
b(\lambda ) &   (a(\lambda^{*})^{*} }
\right],}

where

\eqn\TransMatElems
{a(\lambda )a(\lambda^{*})^{*}+b(\lambda )b(\lambda^{*})^{*}=1.}

The scattering data for the operator $\partial_{-}\sigma$ is
the set
$s=\{ a(\lambda ),b(\lambda );\xi_{j},m_{j},$ $Im(\xi_{j}>0,j=1,...,n\}$.
$a(\lambda )$ can be analytically continued to the half plane 
$Im(\lambda )>0$, $\xi_{j}$ are the zeroes of $a(\lambda )$, and
$m_{j}={-ib(\xi_{j})\over\partial_{-}a(\xi_{j})}$. This last expression
is written with the implicit assumption that we consider only simple
zeroes $\xi_{j}$.

\subsec{Time dependance of the scattering data.}

The transition matrix is independant of $x^{-}$. This allows us to
extract the $x^{+}$ dependance of the transition matrix, in the
limit $x^{-}\to\infty$. Now, recalling
$\lambda = i{g^{2}\over\sqrt{2}}x^{+}+ f(x^{-})$, and using the
boundary condition of the $\sigma$ field, we find

\eqn\First
{{i\sqrt{2}\over g^{2}}\partial_{+}G=x^{-}\sigma_{3}G.}

Alternativly, we may write $G=FT$, and use the boundary condition for $F$,
which implies

\eqn\Second
{{i\sqrt{2}\over g^{2}}\partial_{+}G= x^{-}\sigma_{3}
e^{-i\lambda\sigma_{3}x^{-}}+e^{-i\lambda\sigma_{3}x^{-}}
{i\sqrt{2}\over g^{2}} \partial_{+}T.}

Comparing these last two equations, we find that

\eqn\Third
{{dT\over dx^{+}}=0={\partial T\over\partial x^{+}}+
{\partial\lambda\over\partial x^{+}}
{\partial T\over\partial\lambda}.}

This last equation has the solution

\eqn\Four
{T=T(\lambda -{g^{2}\over\sqrt{2}}x^{+}).}

This is a little unusual. For most models (for example Sine Gordon
model, Heisenberg spin chain, nonlinear Schr\"odinger model) the
spectral parameter is a constant and the scattering evolve in time.
Here we have a linear time dependance for the spectral parameter
and a set of scattering data which do not evolve in time!

\subsec{The inverse scattering problem.}

Takhtajan has shown the kernel $K(x,y;t)$ may be reconstructed from
the time dependant scattering data, using the Gelfand-Levitan-Marchenko
equation\foot{Fogedby\rFog\ has given a clear derivation of this equation
in the context of the classical Heisenberg spin chain.}

\eqn\GLMEqn
{K(x^{-},y^{-};x^{+})+\Phi_{1}(x^{-}+y^{-};x^{+})+\int^{\infty}_{x^{-}}
K(x^{-},z^{-};x^{+})\Phi_{2}(z^{-}+y^{-};x^{+})dz^{-}=0}

where $x^{-}\le y^{-}$,

\eqn\GLMEqnone
{\Phi_{1}=\left[
\matrix{
0 &  -\Lambda^{*}\cr 
\Lambda & 0}
\right],}   

\eqn\GLMEqntwo
{\Phi_{2}=-i\left[
\matrix{
0 &  \partial_{-}\Lambda^{*}\cr 
\partial_{-}\Lambda & 0}
\right]}

and 

\eqn\LambdaDefn
{\Lambda (x^{-};x^{+})={1\over 2\pi}
\int^{\infty}_{-\infty}{b(\lambda ,x^{+})\over\lambda a(\lambda,x^{+})}
e^{i\lambda x^{-}} d\lambda +\sum_{j=1}^{N}
{m_{j}(x^{+})\over\xi_{j}}e^{i\xi_{j}x^{-}}.}

Having solved \GLMEqn\ for a set of scattering data,
$\partial_{-}\sigma$ is then reconstructed using \GPCone\ .

\subsec{Soliton solutions: Reflectionless Potentials.}

If the reflection coefficient $r(\lambda)={b(\lambda)\over a(\lambda)}$
is zero, then the kernel and inhomogeneous terms in the
Gelfand-Levitan-Marchenko equation are reduced to finite sums over
the discrete spectrum. In particular,

\eqn\RefLam
{\Lambda (x^{-};x^{+})=\sum_{j=1}^{N}
{m_{j}(x^{+})\over\xi_{j}}e^{i\xi_{j}x^{-}}.}

Now, making the ansatz

\eqn\AnsatzForK
{K=M(x^{-},x^{+})\left[
\matrix{
\sum_{j}a_{j}e^{i\xi_{j}z^{-}} & 0\cr
0 & \sum_{j} a_{j}^{*}e^{-i\xi_{j}z^{-}}}
\right]}

we find that \GLMEqn\ reduces to a system of linear algebraic equations

\eqn\RefGLMEqn
{\eqalign{M(x^{-},x^{+})&\left[
\matrix{
\sum_{j}a_{j}e^{i\xi_{j}y^{-}} & -\sum_{jk}
{a_{k}m_{j}^{*}e^{-ix_{-}
(\xi^{*}_{j}-\xi_{k})-i\xi^{*}_{j}y^{-}}\over i(\xi_{j}^{*}-\xi_{k})}\cr
-\sum_{jk} {a_{k}^{*}m_{j}e^{ix^{-}(\xi{j}-\xi_{k}^{*})
+i\xi_{j}y^{-}}\over i(\xi_{j}-\xi_{k}^{*})} & \sum_{j}a_{j}^{*}
e^{-i\xi_{j}^{*}y^{-}}}
\right]\cr\noalign {\vskip 0.2truecm}
&+\left[ 
\matrix{
0 & -\sum_{j}{m_{j}^{*}\over\xi_{j}^{*}}e^{-i\xi^{*}_{j}(x^{-}+y^{-})}\cr
\sum_{j}{m_{j}\over\xi_{j}}e^{i\xi_{j}(x^{-}+y^{-})} & 0}
\right]=0. }}

which may be solved explicitly. In this way, we obtain the
$N$ soliton solution for $QCD_{2}$.

For the one soliton solution, we take a single bound state eigenvalue
$\xi$. In this case, the kernel which solves the GLM equation is given by

\eqn\Kernel
{\eqalign{K(x^{-},z^{-})&=
{|m|^{2}(\xi-\xi^{*})^{2}\over |a|^{2}|\xi|^{2}[(\xi-\xi^{*})^{2}
e^{-ix^{-}(\xi-\xi^{*})} -|m|^{2}e^{-ix^{-}(\xi-\xi^{*})}]}\times
\cr\noalign {\vskip 0.2truecm}
&\left[
\matrix{{|a|^{2}\xi\over i(\xi-\xi^{*})}
e^{-i(x^{-}\xi^{*}-\xi z^{-}} & {\xi|a|^{2}\over m}
e^{-i(\xi x^{-}+\xi^{*}z^{-})}\cr
-{|a|^{2}\xi^{*}\over m^{*}} e^{i(\xi^{*}x^{-}+\xi z^{-})} &
{|a|^{2}\xi^{*}\over i(\xi-\xi^{*})} e^{i(x^{-}\xi-\xi^{*}z^{-})}}
\right]
}} 

Inserting this kernel into \GPCone\ yields

\eqn\FirstComponent
{\partial_{-}f_{3}=1-{(2Im(\xi))^{2}\over (2Im(\xi))^{2}
+(2Re(\xi))^{2}} sech^{2}\big[ 2x^{-}Im(\xi)
+ln|{2Im(\xi)\over |m|}|\big],}

and

\eqn\SecondComponent
{\partial_{-}f_{1}+i\partial_{-}f_{2}=\sqrt{1-(\partial_{-}f_{3})^{2}}
e^{i\phi},}

where

\eqn\phiangle
{\eqalign{\phi = arg(m)+2Re(\xi)x^{-}+arctan\Big[ {Im(\xi)\over |\xi|}
\sqrt{|\xi|^{2}\over|\xi|^{2}-(Im(\xi))^{2}}\times
\cr\noalign {\vskip 0.2truecm}
tanh\Big(2x^{-}Im(\xi)+ ln|{2Im(\xi)\over|m|}|\Big)\Big].}}

Now, from the time dependance of the spectral parameter, it is clear that
$m$ is a complex constant, and that

\eqn\SParTime
{\xi ={g^{2}\over\sqrt{2}}x^{+}+\bar{x}^{+}+i{a\over 2},}

where $\bar{x}^{+}$ and $a$ are real constants. Upon inserting this into
\FirstComponent\ and \SecondComponent\ , we find that we are reproducing
the one soliton solution discussed in section $3.2$.

\listrefs
\vfill\eject
\bye